\documentstyle[aps,12pt,preprint,epsfig]{revtex}
 
\def \beq {\begin{equation}}
\def \eeq {\end{equation}}
\def \ba {\begin{eqnarray}}
\def \ea {\end{eqnarray}}

\def \< {\langle}
\def \> {\rangle}
\draft
\begin{document}
 
\title{Polarizations of $J/\psi$ and $\psi'$ in hadroproduction at
Tevatron
 in the $k_t$ factorization approach}
\author{Feng Yuan}
\address{\small {\it Department of Physics, Peking University, Beijing
100871, People's Republic of China\\
Institut f\"ur Theoretische Physik der Universit\"at, Philosophenweg 19, 
D-69120 Heidelberg, Germany}}
\author{Kuang-Ta Chao}
\address{\small {\it China Center of Advanced Science and Technology
(World Laboratory), Beijing 100080, People's Republic of China\\
and Department of Physics, Peking University, Beijing 100871,
People's Republic of China}}
\maketitle
 
\begin{abstract}
We present a calculation for the polarizations
of $J/\psi$ and $\psi'$ produced in the hadron collisions at the Fermilab
Tevatron.
Various color octet channels including ${}^1S_0^{(8)}$, ${}^3P_J^{(8)}$,
and 
${}^3S_1^{(8)}$ as well as contributions from $\chi_{cJ}$ decays are 
considered in 
the $k_t$ factorization
approach. We find that in a rather wide range of the transverse momenta of
$J/\psi$ and $\psi'$, the production rates could be dominated by the 
${}^1S_0^{(8)}$ channel, and the predicted
polarizations from the ${}^1S_0^{(8)}$ channel and $\chi_{cJ}$ 
feeddown contributions
are roughly compatible with the preliminary CDF data.
This might provide a possible release from the conflict between the
NRQCD collinear parton model calculations and the CDF data.
\end{abstract}
 
Heavy quarkonium production in high energy collisions provide
important information on both perturbative and nonperturbative QCD.
In recent years, heavy quarkonium production has attracted much attention
from both theory and experiment.
To explain the $J/\psi$ and $\psi'$ surplus problem
of large transverse momentum production at the Fermilab Tevatron\cite{fa},
the color-octet production mechanism was introduced for the
description of heavy quarkonium production\cite{surplus} based
on the NRQCD factorization framework\cite{nrqcd}.
However, most recently the CDF collaboration has reported 
preliminary measurements on the polarizations of the promptly
produced charmonium states\cite{pola}, which appear not to
support the color-octet predictions that the directly produced $S$-wave
quarkonia have transverse polarizations at large $p_T$
\cite{th-pola,beneke}.
In \cite{braaten}, the authors further
considered the feeddown contributions from
$\chi_c$ decays, and found the prompt $J/\psi$ polarization disagree with 
the CDF data at large $p_T$ by 3 standard deviations.
In \cite{ktf1}, we have performed a calculation of the production rates
for
$J/\psi$ and $\psi'$ at Tevatron in the newly advocated $k_t$
factorization 
approach\cite{teryaev}, and we find that within the color-singlet model 
the production rates of $J/\psi$ and $\psi'$ can be
enhanced by a factor of 20 compared to the leading order predictions 
in the naive collinear parton
model, but are still below the data by at least one order of magnitude.
Therefore in order to explain the $J/\psi$ and $\psi'$ productions at 
Tevatron we still need to call for the contributions from color-octet 
channels. 
This conclusion has been confirmed by another independent study on the
$J/\psi$ production in the $k_t$ factorization approach\cite{t1}.
In this letter we will report a calculation for 
 the contributions of color-octet 
channels not only to the cross sections but also to the polarizations 
of the $J/\psi$ and $\psi'$.
 
The $k_t$-factorization approach differs greatly from the
conventional collinear approximation because it takes the non-vanishing
transverse momenta of the scattering partons into account.
This approach was previously advocated for the study of heavy quark
production
at high energy hadron colliders\cite{ca,co}.
Recently, more phenomenological studies were performed along this 
direction by several
groups\cite{ryskin,zotov,hagler}, and some successes seem to be achieved
by 
comparing the calculations with experimental
data on heavy quark production at the DESY HERA and the Fermilab Tevatron 
\cite{zotov,hagler,schleper}. Especially
for $b$ and $\bar b$ production, the $k_t$ factorization approach may
provides a successful resolution for the long standing conflict between
the NLO collinear parton model prediction and the experimental data 
at the Tevatron \cite{hagler}.
From these and the most recent studies on charmonium production
\cite{ktf1,teryaev,t1},
we see that for heavy quark and quarkonium 
production in the present $p_T$ region at the Tevatron, 
the improved $k_t$ factorization approach may play a very important role
and bring significant enhancements 
compared with the
conventional collinear parton model. For the $b$ quark production the 
enhancement is about a factor of $2$, while for the color-singlet $J/\psi$ 
production a more dramatical  
enhancement, which is about a factor of $20$, has been found \cite{ktf1}.
In this paper we will
work along this line, to further study $J/\psi$ polarization at the
Tevatron
by including the color-octet channels.
We find that the $k_t$ factorization approach may provide a possible way
to understand both the observed production rates and polarizations of 
$J/\psi$ and $\psi'$.
 
Under the $k_t$ factorization approach, a process can be
factorized as the hard scattering cross section convoluted with the
so called unintegrated gluon distribution of the proton. For example, 
the polarized cross section for a charmonium state $H$ at the Tevatron 
can be written as the following form,
\begin{equation}
\label{xs}
\frac{d\sigma(p\bar p\rightarrow H^{(\lambda)} X)}{d^2P_Tdy}=\int dx_1dx_2
d^2q_{1T}d^2q_{2T}\frac{f(x_1;q_{1T}^2)}{q_{1T}^2}
\frac{f(x_2;q_{2T}^2)}{q_{2T}^2}\times \hat \sigma (g_1^*g_2^*\rightarrow 
H^{(\lambda)}+X),
\end{equation}
where $P_T$ and $y$ are the transverse momentum and rapidity of the
charmonium
state $H$, and $\lambda$ denotes its helicity.
$f(x;q_{T}^2)$ is the unintegrated gluon distribution function,
which can be related to the conventional gluon distribution by
\begin{equation}
xg(x,\mu^2)=\int^{\mu^2} \frac{dq_T^2}{q_T^2}f(x;q_T^2).
\end{equation}
In Eq.~(1), the hard scattering cross section
$\hat \sigma(H^{(\lambda)})$ describes the production of $H$ 
from two off-shell gluons fusion process, 
which will depend on the incident gluons' longitudinal momentum fractions
$x_1,~x_2$ and transverse momenta $q_{1T},~q_{2T}$.
In the collinear approximation, the cross section $\hat \sigma$ does not 
depend on 
$q_{1T}$ and $q_{2T}$, and then Eq.~(1) will go back to the conventional 
collinear parton model result for the production of state $H$. 
However, as mentioned above, the incident partons' $q_t$ may play 
an important
role for heavy quarkonium production, and may not be neglected.
To derive the cross section $\hat \sigma$, we closely follow the 
BFKL approach in which gauge invariance of the interaction 
vertices is guaranteed order by order. 
For example, to get the gauge invariant amplitude for $J/\psi$ production
from ${}^3S_1^{(8)}$ channel, we must include extra term contributions
(see Fig.~1(c) of \cite{ktf1})
which will finally result in an effective (Lipatov) vertex for three-gluon 
interaction.
 
For direct $J/\psi$ production at the Tevatron, because the color-singlet
contribution underestimates the experimental data\cite{ktf1} 
by at least an order of magnitude
in the $k_t$ factorization approach,  
we need to consider the color-octet 
contributions, for which we will have the 
LO (leading order) gluon fusion processes,
\begin{equation}
gg\rightarrow c\bar c[{}^1S_0^{(8)},{}^3P_J^{(8)},{}^3S_1^{(8)}]
\rightarrow J/\psi+X,
\end{equation}
where various possible color-octet channels for the $c\bar c$ are
considered.
Note that these $2\rightarrow 1$ subprocesses do not contribute 
to $J/\psi$ ($\psi'$)
production at large $p_T$ in the collinear parton model. However, in the 
$k_t$ factorization approach, these processes contribute to large $p_T$ 
production, because the initial partons' transverse momenta will 
lead to $J/\psi$ having transverse momentum. 
In this context, the transverse momentum of $J/\psi$ will be the sum of
the transverse momenta of the two incident partons,
$\vec{P}_T^\psi=\vec{q}_{1T}+\vec{q}_{2T}$.
For $\chi_{cJ}$ production, the inclusive production rates have been
calculated in \cite{teryaev}, where they find there is no need for the 
color-octet 
contributions to $\chi_{cJ}$ production. So, in this paper we will also
only
consider the color-singlet contributions to $\chi_{cJ}$ polarizations,
and then calculate their feeddown contributions to $J/\psi$ polarization.
For numerical calculations, we set $m_c=1.5GeV$ and 
choose the unintegrated gluon distribution
of \cite{martin}. We set the scales $\mu^2$ for
the strong coupling constant $\alpha_s(\mu^2)$ in the
hard scattering cross section $\hat \sigma$
to be $q_{1T}^2$ for the interaction
vertex associated with the incident gluon $q_1$, and 
$q_{2T}^2$ for the vertex associated with $q_2$\cite{ktf1,teryaev,levin}.
 
We first give the numerical results for the $P_T$ distributions of the 
inclusive cross sections for direct $J/\psi$ and $\psi'$ production at
the Tevatron compared with the experimental data. 
Fig.~1 is for $J/\psi$, and Fig.~2 for $\psi'$.
As a qualitative estimate we will use just one individual color-octet 
channel to fit the cross sections observed by CDF.    
The solid lines correspond to the ${}^1S_0^{(8)}$ channel by taking the
color-octet matrix elements as:
$\langle {\cal O}_8^{\psi}({}^1S_0)\rangle =0.10GeV^3$ for $J/\psi$ and 
$\langle {\cal O}_8^{\psi'}({}^1S_0)\rangle =0.03GeV^3$ for $\psi '$,
while
other matrix elements are set to be zero,
$\langle {\cal O}_8^{\psi(\psi')}({}^3P_0)\rangle =
\langle {\cal O}_8^{\psi(\psi')}({}^3S_1)\rangle =0$.
The dashed lines correspond to the ${}^3P_J^{(8)}$ channel:
$\langle {\cal O}_8^{\psi}({}^3P_0)\rangle /m_c^2 =0.025GeV^3$ for $J/\psi$ and 
$\langle {\cal O}_8^{\psi'}({}^3P_0)\rangle /m_c^2=0.005GeV^3$ for $\psi '$,
while
$\langle {\cal O}_8^{\psi(\psi')}({}^1S_0)\rangle =
\langle {\cal O}_8^{\psi(\psi')}({}^3S_1)\rangle =0$.
The dotted lines correspond to the ${}^3S_1^{(8)}$ channel:
$\langle {\cal O}_8^{\psi}({}^3S_1)\rangle =0.007GeV^3$ for $J/\psi$ and 
$\langle {\cal O}_8^{\psi'}({}^3S_1)\rangle =0.0025GeV^3$ for $\psi '$,
while
$\langle {\cal O}_8^{\psi(\psi')}({}^1S_0)\rangle =
\langle {\cal O}_8^{\psi(\psi')}({}^3P_0)\rangle =0$.
For $J/\psi$ production, from Fig.~3 we can see that the ${}^1S_0^{(8)}$
and 
${}^3P_J^{(8)}$ channels can both rather well describe the Tevatron data, 
especially for the shapes of the differential cross sections.
Whereas the ${}^3S_1^{(8)}$ channel gives an inadequate shape for the 
differential
cross sections compared with the experimental data.
For $\psi'$ production, at first sight all the three curves seem to 
roughly agree with data.
However, we can still see that the ${}^3S_1^{(8)}$ channel 
overestimates the production rate of $\psi'$ at large $P_T$.
For both $J/\psi$ and $\psi'$ the theoretical predictions 
for ${}^3S_1^{(8)}$ channel drop more
slowly as $P_T$ increases than the experimental data.
This is mainly because for  ${}^3S_1^{(8)}$ channel there are additional
terms (see Fig.~1b and c of \cite{ktf1}), which give main contributions 
at large $P_T$, but these terms do not 
contribute to the ${}^1S_0^{(8)}$ and ${}^3P_J^{(8)}$ processes.
The disagreements between theoretical predictions for the ${}^3S_1^{(8)}$ 
contributions and experimental data show that 
in the present $p_T$ region ($p_T<20GeV$) at the Tevatron, the
${}^3S_1^{(8)}$ 
channel may not be the dominant one to the direct $J/\psi$ and $\psi'$ 
production in the $k_t$ factorization approach.
This result is very different from the collinear parton model 
calculations,
where the ${}^3S_1^{(8)}$ channel overwhelmingly dominates the $S$-wave 
quarkonium production at moderate and large transverse momentum at the
Tevatron\cite{surplus,beneke}.
 
This difference has significant consequence on the polarization
predictions
of $J/\psi$ and $\psi'$. We recall that in the 
collinear parton model the NRQCD formalism predicts $S$-wave quarkonia 
being transversely polarized dominantly due to the  ${}^3S_1^{(8)}$ 
gluon fragmentation contributions\cite{th-pola,beneke}.
However, in the $k_t$ factorization approach,  the ${}^3S_1^{(8)}$ 
channel may not dominate the $S$-wave quarkonium production, so the
previous
conclusion of transverse polarization at large $P_T$ may be changed.
In the following we will study the polarizations of $J/\psi$ and $\psi'$
in the $k_t$ factorization approach.
In Fig.~3, we present our result for the $\psi'$ polarization at Tevatron.
Because it receives no feeddown contributions from higher lying 
charmonium states,
the polarization of $\psi'$ is only due to the direct production 
from the color-octet processes.
In this figure, we give each individual contributions from these octet 
processes in the $k_t$ factorization approach. 
The solid line is for the ${}^1S_0^{(8)}$ 
contribution, the dashed line for the ${}^3P_J^{(8)}$,
and the dotted-dashed line for the ${}^3S_1^{(8)}$. 
From this figure we can see that the ${}^3S_1^{(8)}$ channel contributes
to the transversely polarized $\psi'$ as it contributes in the collinear
parton model approach.
Whereas the ${}^1S_0^{(8)}$ channel contributes to the unpolarized, and  
${}^3P_J^{(8)}$ contributes to the longitudinally polarized.
If  ${}^1S_0^{(8)}$ could dominate $\psi'$ production
at the Tevatron (see Fig.~2), the $\psi'$ will be unpolarized. 
This is consistent with the experimental data, which show that 
in the available $p_T$ region the $\psi'$ seems to be unpolarized (though
with 
large errors).
So, the experimental data on $\psi'$ polarization support the 
${}^1S_0^{(8)}$ dominance on $\psi'$ production.
 
We then study the $J/\psi$ polarization. The polarization data
of $J/\psi$ are for the prompt production, i.e., including the direct
production and the feeddown contributions from $\chi_{cJ}$ and $\psi'$
decays. The $\psi'$ feeddown contributions to $J/\psi$ polarization can be
obtained from the above $\psi'$ polarization calculations multiplied by
the
decay branching ratio of $\psi'$ to $J/\psi$.
(Note that the observed transitions of $\psi'$ to $J/\psi$ involve no spin 
flips, so this part of contribution has the same behavior as $\psi'$ 
polarization.)
To obtain the feeddown contributions from the $\chi_{cJ}$ decays, we
must convolute the $\chi_{cJ}$ polarized cross sections with their 
decay branching ratios of $\chi_{cJ}\rightarrow J/\psi \gamma$.
We find that in the $k_t$ factorization approach $\chi_{cJ}$ feeddown
contributes to $J/\psi$ being a little transversely polarized.
The polarization parameter $\alpha$ for $J/\psi$ from  $\chi_{cJ}$ decays 
approaches to $0.5$ at large $p_T$.
This is different from the collinear parton model calculations
which predict that $\alpha$ approaches to about $0.24$ at large
$p_T$\cite{wise,braaten}.
(Note that the polarizations of $\chi_{cJ}$ themselves are 
very interesting for testing the
$k_t$ factorization approach, and has been presented in\cite{ktf2}).
All of these contributions including the direct contributions and the
feeddown 
contributions to the $J/\psi$ polarization are plotted in Fig.~4.
As in Fig.~3 for $\psi'$, we also plot in this figure the individual 
contributions 
from the three color-octet processes for the direct polarizations, 
and then combine them
with the $\chi_{cJ}$ and $\psi'$ feeddown contributions for the prompt
polarizations
compared with the experimental data from the CDF collaboration.
The thinner lines are for the direct polarizations, and the thicker 
lines for the prompt polarizations.
The solid lines are for ${}^1S_0^{(8)}$ contributions,
the dashed lines for ${}^3P_J^{(8)}$ , the dotted-dashed
lines for ${}^3S_1^{(8)}$, and the dotted line for the $\chi_{cJ}$ 
feeddown contribution.
We see that in general the polarizations increase as $p_T$ increases.
For example, for the ${}^3P_J^{(8)}$ direct production the value of 
polarization
parameter $\alpha(\psi)$ (being longitudinal) varies from
$-0.63$ at $p_T=5GeV$ to $-0.38$ at $p_T=20GeV$. For the ${}^3S_1^{(8)}$ 
direct production  $\alpha(\psi)$ (being transverse) varies from $0.72$
to $0.96$ in the same $p_T$ range. The $\chi_c$ feed down contribution
will
reduce the extremities of $\alpha(\psi)$ in direct productions.   
From this figure we find again that the ${}^1S_0^{(8)}$ dominance
together with the $\chi_c$ feed down contribution will lead to the 
predicted prompt $J/\psi$ polarization in rough 
agreement with the Tevatron data except the one at about $p_T=16.7 GeV$
which, however, has very large errors.
We note that this seemingly nontrivial behavior of $J/\psi$ polarization 
at $p_T=16.7 GeV$ (being longitudinal)
can not be explained consistently with the moderate $p_T$ data
(being transverse) in our approach.
If the CDF data at $p_T=16.7GeV$ is further confirmed by
CDF or D0 experiments in the future with higher statistics, 
the above mechanism may fail. 
 
Concluding our analysis, we find that the ${}^1S_0^{(8)}$ channel 
dominance is the only reasonable realization of the polarization
data of $J/\psi$ and $\psi'$ in the $k_T$ factorization approach.
To see this point more clearly, we can perform a $\chi^2$ analysis
for theoretical comparison with the experimental data.
For $\psi'$ polarization data, ${}^1S_0^{(8)}$ channel gives the smallest
value
of $\chi^2$ ($\chi^2/dof=0.61$) as expected.
Moreover, inclusion of other channels
does not improve the fit. For example, including the 
${}^3S_1^{(8)}$ contribution would make the fit even worse, and when
including ${}^3P_J^{(8)}$ contribution the fit is in the same
manner as ${}^1S_0^{(8)}$ alone but with a little 
higher $\chi^2/dof$ ($=0.83$). 
For $J/\psi$ polarization, the result for the fit is similar. I.e.,
${}^1S_0^{(8)}$ alone gives the smallest $\chi^2$ value
($\chi^2/dof=1.85$), and inclusion of ${}^3P_J^{(8)}$ gives a 
even worse fit, while inclusion of ${}^3S_1^{(8)}$ gives the same quality
fit as ${}^1S_0^{(8)}$ alone with 
a little higher $\chi^2/dof$ ($=2.13$).
So, with the available statistics at present, 
the polarization data of $J/\psi$
and $\psi'$ may indicate a strong signal of ${}^1S_0^{(8)}$ 
dominance for $J/\psi$ and $\psi'$ production.
 
The ${}^1S_0^{(8)}$ dominance has a straightforward consequence
that the other two matrix elements $\langle {\cal
O}_8^{\psi}({}^3S_1)\rangle$
and $\langle {\cal O}_8^{\psi}({}^3P_0)\rangle$ must be
much smaller than $\langle {\cal O}_8^{\psi}({}^1S_0)\rangle$.
This is quite different from the naive expectation based on
the NRQCD velocity scaling rules, which predict no big difference 
between these three matrix elements 
($\langle {\cal O}_8^{\psi}({}^1S_0)\rangle$ scales as $v^6$, while 
$\langle {\cal O}_8^{\psi}({}^3S_1)\rangle$
and $\langle {\cal O}_8^{\psi}({}^3P_0)\rangle$ both scale as $v^7$).
But, the values of
$\langle {\cal O}_8^{\psi(\psi')}({}^1S_0)\rangle$ we used in 
Figs.~1-2 are smaller than the relevant color-singlet matrix elements
by about an order of magnitude (e.g., the observed $J/\psi$ leptonic
decay width leading to
$\langle {\cal O}_1^{\psi}({}^3S_1)\rangle=1.06GeV^3$).
This is roughly consistent with the NRQCD velocity scaling rule 
estimate. As for the other two matrix elements 
$\langle {\cal O}_8^{\psi}({}^3S_1)\rangle$
and $\langle {\cal O}_8^{\psi}({}^3P_0)\rangle$,
there might be some dynamical reasons (associated with the nonperturbative
evolution processes) to suppress their values and violate the
velocity scaling rules\cite{wong}, or there might be some 
new counting rules for
charmonium system to suppress their values\cite{fleming}.
 
Some theoretical uncertainties must be addressed before concluding.
First there is an uncertainty coming from the scale choice used for the
numerical calculations. We have set the scales to be $q_{1T}^2$ or
$q_{2T}^2$
which are also adopted by some previous studies
\cite{hagler,levin}.
If we set these scales both equal to $p_T^2(\psi)+m_c^2$, the cross
sections
for the three color-octet channels will be lowered down 
with the normalization factor changing from $1$ to about $0.6$
both for $J/\psi$ and $\psi'$. 
But this change has little effects on the shapes of the 
curves in Figs.~1-2. So, our conclusion for the ${}^1S_0^{(8)}$ dominance
will not change
with the scale changes.
Another uncertainty may come from the
parameterizations of the unintegrated gluon distributions, for which,
however, it is
difficult to give an estimate because we 
have little knowledge about it at present. 
On the other hand, we note
that the parameterization provided by \cite{martin} are obtained from an
excellent fit to $F_2(x,Q^2)$ in a very large window of $x$ and $Q^2$
in the $k_t$ factorization approach,
which may be viewed as a reliable determination of these quantities.
 
In conclusion, in this paper we have calculated the $J/\psi$ and $\psi'$
production rates and polarizations in hadroproduction at the Tevatron 
in the $k_t$ factorization approach.
Various color-octet contributions including 
${}^1S_0^{(8)}$, ${}^3P_J^{(8)}$, and 
${}^3S_1^{(8)}$ as well as contributions from $\chi_{cJ}$ decays are
considered 
in this approach. 
We find that in a rather wide range of the transverse momenta of
$J/\psi$ and $\psi'$, the production rates could be dominated by the 
${}^1S_0^{(8)}$ channel, and the predicted
polarizations are in rough agreement with the preliminary Tevatron
data.
However, to draw the final conclusion on heavy quarkonium 
production mechanism
in high energy hadron collisions, 
we still need more accurate experimental data on large $p_T$
production cross sections and polarizations.
We also need more solid theoretical 
calculations such as NLO corrections to the color-octet processes
discussed 
above, and a more reliable estimate 
for the so-called unintegrated gluon distribution function. It is
worthwhile
to make further investigations along this direction.
 
While the calculation in this paper was finished we noticed
a paper by Ph.~H\"aegler et al. \cite{t1},
 who studied the $J/\psi$ production rate in the
$k_t$ factorization approach with color octet channels. Their result is
similar to ours for the calculation of cross sections.
 
\acknowledgments
This work was supported in part by the National Natural Science Foundation
of China, the Education Ministry of China, and the State
Commission of Science and Technology of China.

\newpage
\centerline{\bf \large Figure Captions}
\vskip 1cm
\noindent
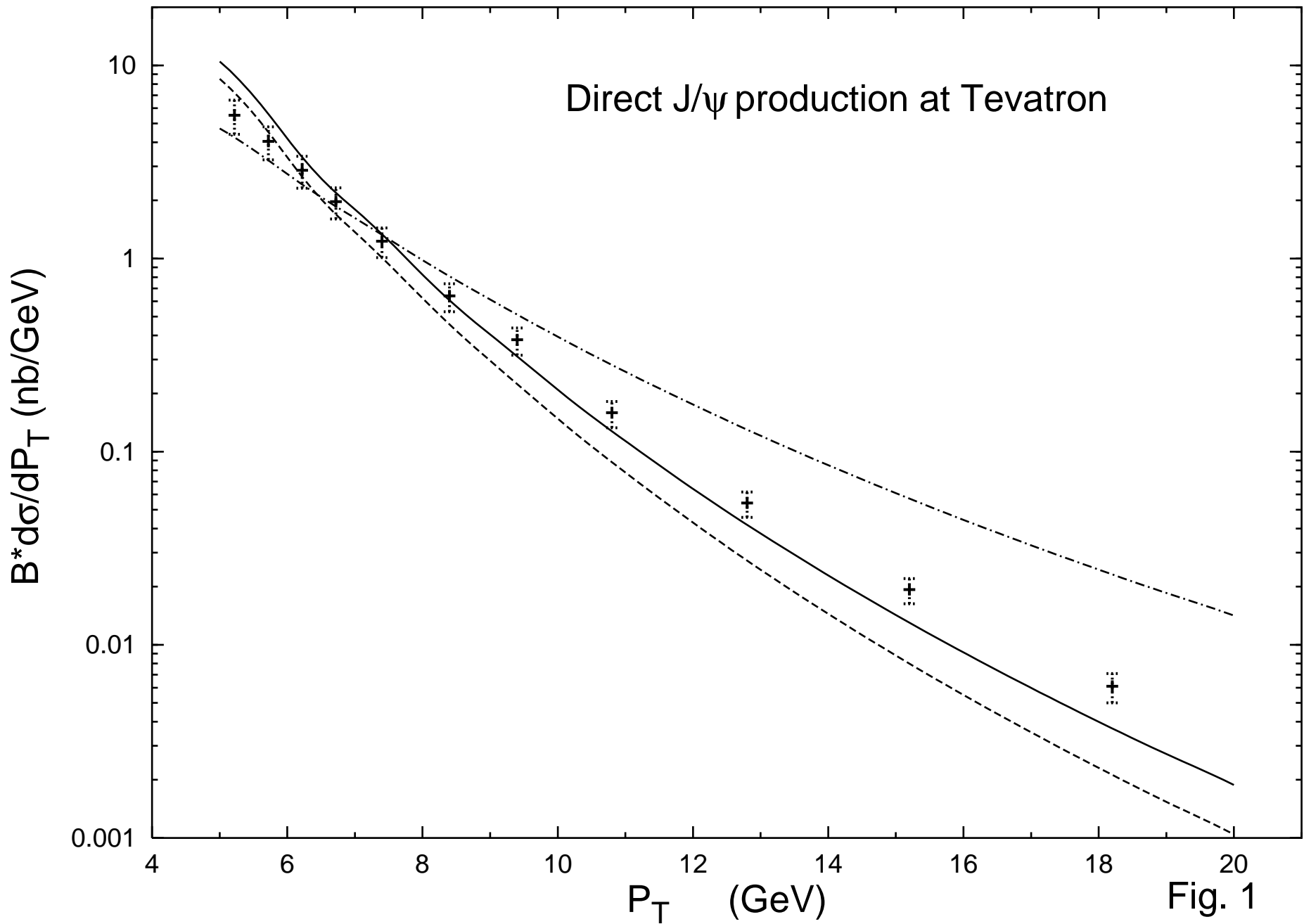
FIG. 1. The $p_T$ distribution of direct $J/\psi$ production cross section
at the Tevatron in the $k_t$ factorization approach. The solid line is for 
${}^1S_0^{(8)}$ contribution, the dashed line for ${}^3P_J^{(8)}$, and
the dotted-dashed line for ${}^3S_1^{(8)}$.
 
\noindent
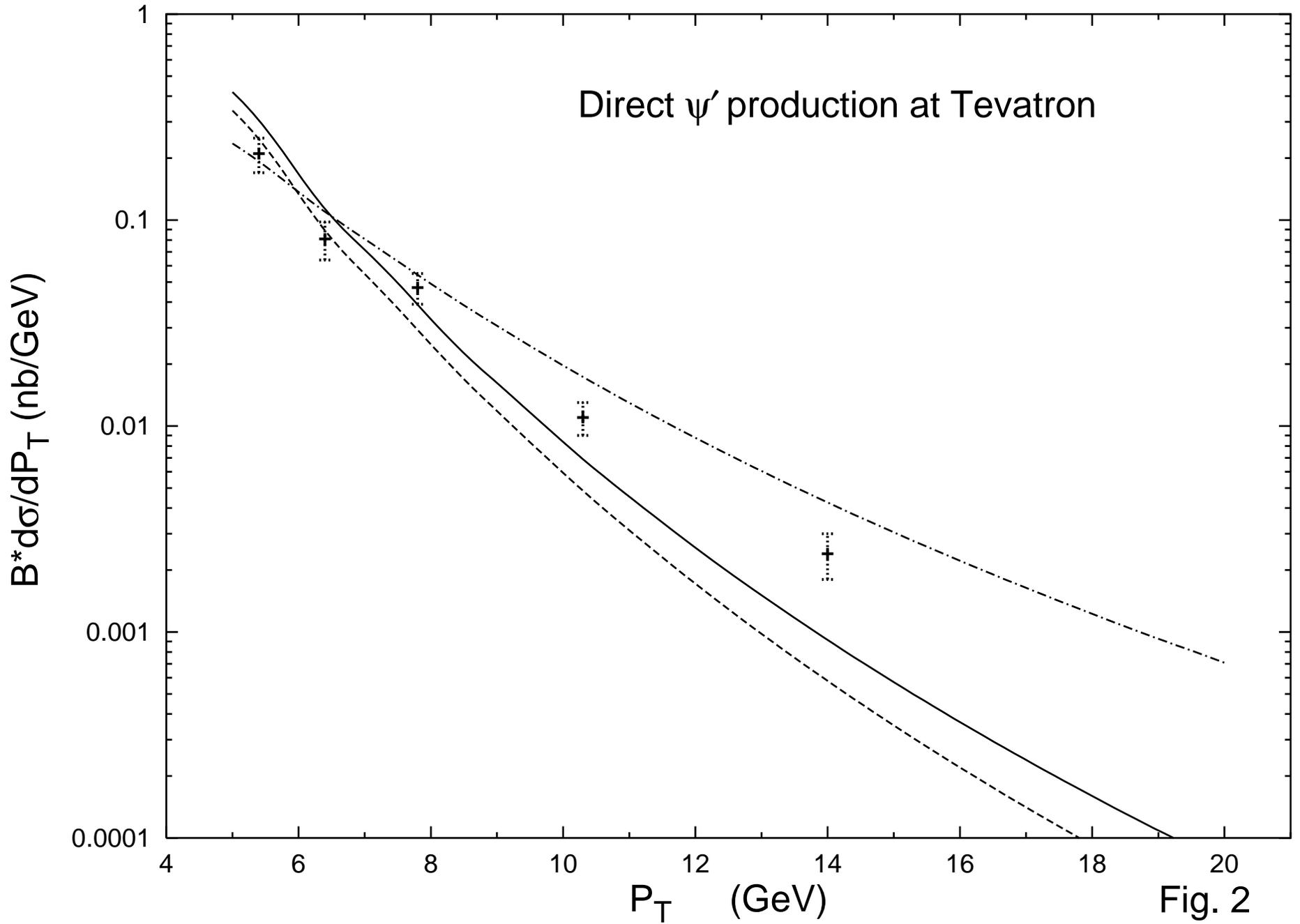
FIG. 2. The same as Fig.~1 but for $\psi'$.
 
\noindent
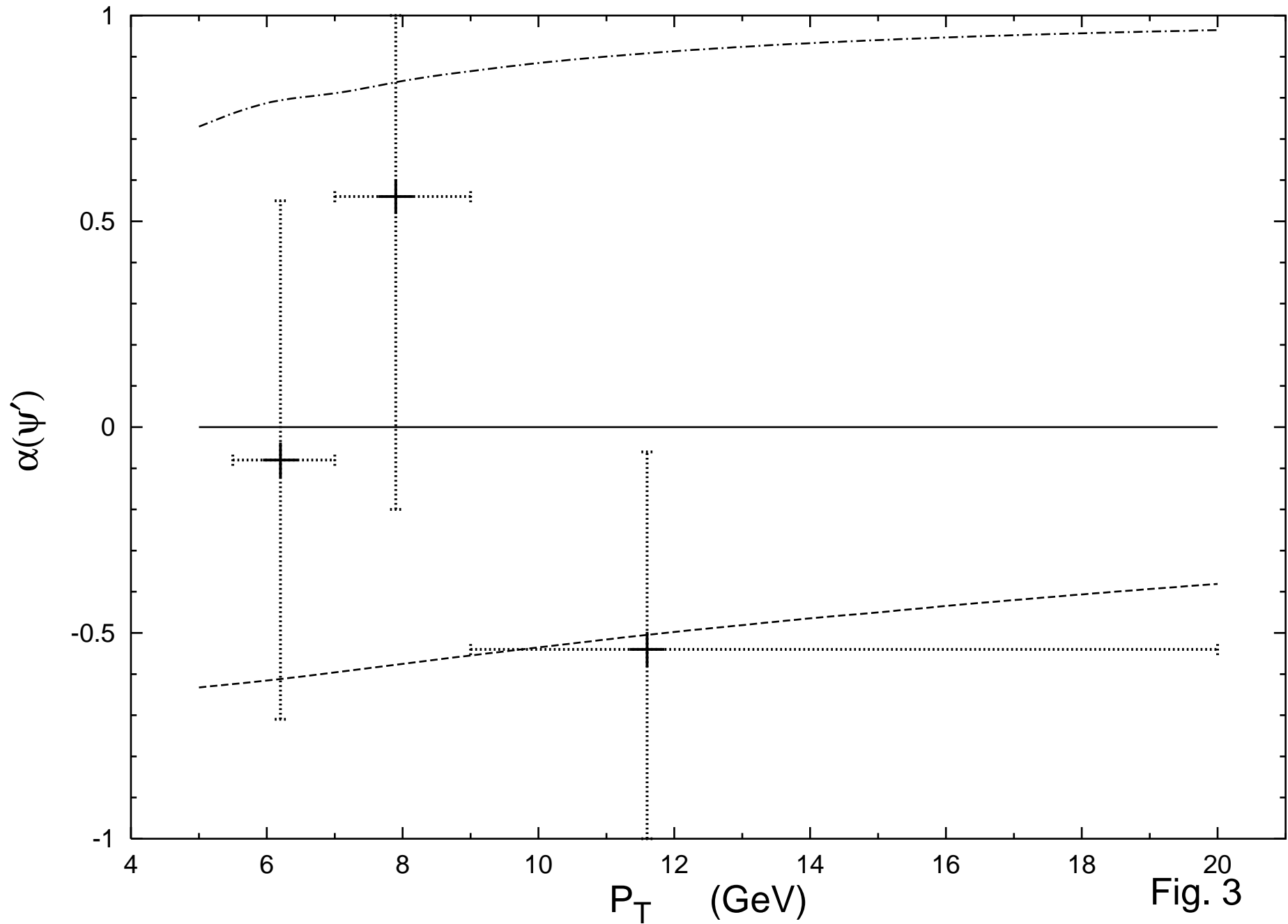
FIG. 3. Prompt $\psi'$ polarization at the Tevatron.

\noindent
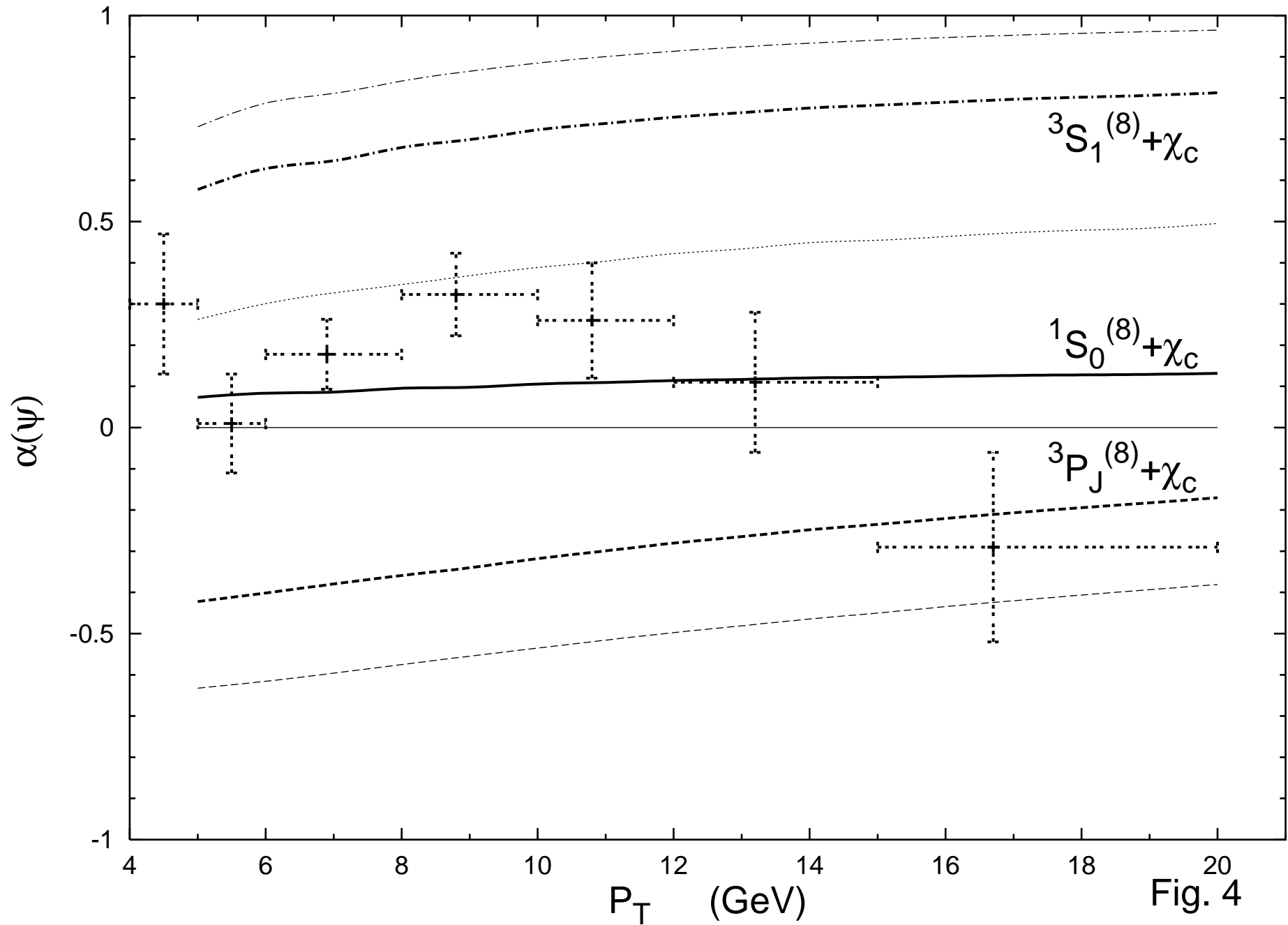
FIG. 4. Prompt $J/\psi$ polarization at the Tevatron.
The thinner lines are for the direct polarizations, and the thicker lines 
for the prompt polarizations.

\end{document}